\documentclass[12pt]{article}
\global\arraycolsep=2pt 
\input{epsf}
\begin{document}
\makeatletter
\def\fmslash{\@ifnextchar[{\fmsl@sh}{\fmsl@sh[0mu]}}
\def\fmsl@sh[#1]#2{%
  \mathchoice
    {\@fmsl@sh\displaystyle{#1}{#2}}%
    {\@fmsl@sh\textstyle{#1}{#2}}%
    {\@fmsl@sh\scriptstyle{#1}{#2}}%
    {\@fmsl@sh\scriptscriptstyle{#1}{#2}}}
\def\@fmsl@sh#1#2#3{\m@th\ooalign{$\hfil#1\mkern#2/\hfil$\crcr$#1#3$}}
\makeatother
\thispagestyle{empty}
\begin{titlepage}
\begin{flushright}
hep-ph/0008243 \\
LMU 10/00 \\
\today
\end{flushright}

\vspace{0.3cm}
\boldmath
\begin{center}
\Large {\bf The Electroweak Interactions as a Confinement Phenomenon.}
\end{center}
\unboldmath
\vspace{0.8cm}
\begin{center}
  {\large Xavier Calmet}\\
  
\end{center}
\begin{center}
  and
  \end{center}
\begin{center}
{\large Harald Fritzsch}\\
 \end{center}
 \vspace{.3cm}
\begin{center}
{\sl Ludwig-Maximilians-University Munich, Sektion Physik}\\
{\sl Theresienstra{\ss}e 37, D-80333 Munich, Germany}
\end{center}
\vspace{\fill}
\begin{abstract}
\noindent
We consider a model for the electroweak interactions based on the
assumption that physical particles are singlets under the gauge group
$SU(2)$. The concept of complementarity explains why the standard
model works with such an extraordinary precision although the fermions
and bosons of the model can be viewed as composite objects of some
more fundamental fermions and bosons. We study the incorporation of
QED in the model. Furthermore we consider possible deviations from the
standard model at very high energies, e.g. excited states of the weak
bosons.
\end{abstract}
to appear in Physics Letters B 
\end{titlepage}
\section{Introduction}
The standard model of the basic interactions consists of two sectors,
the QCD-sector based on an unbroken and confining gauge theory in color
space, and the electroweak sector \cite{Glashow}, based on the gauge group
$SU(2)_L \otimes U(1)_Y$, which is spontaneously broken. In QCD the
gauge bosons (gluons) are massless but three of the gauge bosons of
the electroweak sector acquire masses through the spontaneous symmetry
breaking. This looks like a peculiar asymmetry between the strong and
electroweak sectors, and many authors have attempted to avoid it by
extending the standard model, e.g. by considering composite models
\cite{Fritzsch} for leptons, quarks and the weak bosons.

As in particular emphasized by 't Hooft, the asymmetry between the two
sectors is, in fact, much less pronounced if one views the electroweak
gauge model from a different point of view \cite{'tHooft:1998pk},
using the idea of complementarity
\cite{Osterwalder:1978pc,Fradkin:1979dv}. Like in QCD, the electroweak
sector can be built upon a confining gauge theory. But in contrast to
QCD such a theory does not show a clear-cut distinction between the
confining phase and the Higgs phase, like there is no such distinction
between the gaseous and liquid phases of water. A continuous
transition between the two phases is possible.

In this paper we take this change of viewpoint seriously and suggest
that the observed electroweak interactions are indeed due to a
confining gauge theory. As long as the confinement scale is
disregarded, the theory is identical to the standard model of the
electroweak interactions as emphasized in ref. \cite{'tHooft:1998pk}.
But we shall demonstrate that the observations, in particular the
structure of the neutral current interaction and the strength of the
electroweak mixing angle $\theta_W$, suggest that new non-perturbative
effects (e.g.  bound state effects) might show up at the energy scale
of the order of several hundred GeV.

\section{The electroweak gauge group and confinement} \label{sec2}

We start by writing down the Lagrangian of the electroweak standard
model, taking into account only the first family of leptons and quarks:
\begin{eqnarray} \label{SML}
  {\cal L}_{h}&=&-\frac{1}{4} F^{a}_{\mu \nu}  F^{a \mu \nu}
                 -\frac{1}{4} f_{\mu \nu}  f^{ \mu \nu}
                 + \bar L_L i \fmslash{D} L_L+ \bar Q_L i \fmslash{D} Q_L
                 +\bar e_R i \fmslash{D} e_R\\ \nonumber &&
                 +\bar u_R i \fmslash{D} u_R
                 + \bar d_R i \fmslash{D} d_R+G_e \bar e_R (\tilde{\phi} L_L)
                 +G_d \bar d_R (\tilde{\phi} Q_L) \\ \nonumber &&
                 +G_u \bar u_R (\phi Q_L)
                 +h.c.
                 +\frac{1}{2}(D_{\mu} \phi)^\dagger (D^{\mu} \phi)
                 -\frac{\mu^2}{2} \phi \phi^\dagger
                 -\frac{\lambda}{4} (\phi \phi^\dagger)^2.
\end{eqnarray}
The covariant derivative is given by:
\begin{eqnarray}
        D_{\mu}=\partial_{\mu}-i \frac{g'}{2} Y {\cal A}_\mu -
        i \frac{g}{2} \tau^a B^a_{\mu}. 
\end{eqnarray}
The field strength tensors are as usual
\begin{eqnarray}
F^a_{\mu \nu}&=&\partial_\mu B^a_\nu-  \partial_\nu B^a_\mu
+g \epsilon^{abc} B^b_\mu B^c_\nu \\
f_{\mu \nu}&=&\partial_\mu {\cal A}_\nu-  \partial_\nu {\cal A}_\mu.
\end{eqnarray}
We have used the definitions:
\begin{eqnarray}
L_L=\left(\begin{array}{c} \nu_L \\ e_L \end{array}
\right ),  Q_L= \left(\begin{array}{c} u_L \\ d_L \end{array}
\right ), \phi=\left(\begin{array}{c} \phi^0 \\ \phi^- \end{array}
\right ) \mbox{and} \,
\tilde{\phi}=i \sigma_2 \phi^*=\left(\begin{array}{c} \phi^+ \\ - {\phi^0}^* \end{array}
\right ).
\end{eqnarray}

According to the well-known mechanism of spontaneous symmetry breaking
the neutral component of the scalar field $\phi^0$ acquires a
non-vanishing vacuum expectation value $v$, which gives rise in
particular to the masses of the electroweak gauge bosons.

The Lagrangian (\ref{SML}) admits a much wider class of solutions, if
we vary the parameters $\mu^2$ and $\lambda$ arbitrarily. If $\mu^2$
is positive, the $SU(2)_L$ gauge bosons remain massless and lead to a
confinement of the corresponding electroweak charges. As pointed out
by Osterwalder and Seiler \cite{Osterwalder:1978pc} and, using the
lattice approach, by Fradkin and Shenker \cite{Fradkin:1979dv}, there
is no phase transition between the Higgs phase and the confinement
phase if the theory has a scalar field in the fundamental
representation of the gauge group.  However, they were restricting
themselves to the case of an $SU(2)$ gauge group without fermions and
using the so-called frozen Higgs approximation.

The absence of a phase transition allows one to carry out a smooth
transition between the Higgs phase and the confinement phase, a
phenomenon denoted as complementarity. However some care has to be
taken with the notion of complementarity since it was shown by
Damgaard and Heller \cite{Damgaard:1985nb}, performing a mean field
analysis, that for certain values of the parameters a phase transition
can appear. We assume that the parameters in reality are such that the
concept of complementarity can be applied.

Using this phenomenon we can start out from the electroweak Lagrangian
(\ref{SML}) and study the confinement phase. In particular we shall
concentrate on possible departures from the electroweak standard model
at high energies and on the incorporation of the $U(1)_Y$ hypercharge
which leads to the electromagnetic interaction.

In the confinement phase the physical particles are $SU(2)_L$
singlets. We introduce the following fundamental left-handed
dual-quark doublets, which we denote as D-quarks:
\\
\begin{tabular}{lll}
leptonic D-quarks & $l_i=  \left(\begin{array}{c} l_1 \\ l_2 \end{array}

\right )$  &  (spin   $1/2$,  left-handed)  \\
& & \\
 hadronic  D-quarks   & $q_i= \left(\begin{array}{c}q_1 \\ q_2\end{array}
\right )$   &  (spin $1/2$, left-handed, $SU(3)_c$ triplet) \\
& & \\
scalar D-quarks  &  $h_i= \left(
  \begin{array}{c}
  h_1 \\ h_2
  \end{array}
\right )$ & (spin $0$). \\  \\
\end{tabular}
\\
The right-handed particles are those of the standard model. We can
identify the left-handed fermions and electroweak bosons of the
standard model as bound states:

\begin{eqnarray} \label{def1}
 &{\rm neutrino}: &  \nu_L \propto \bar h l  \nonumber  \\ 
 &{\rm electron}: &  e_L\propto h l  \nonumber \\
 &{\rm up \ type \ quark}:&  u_L\propto \bar h q  \nonumber \\
 &{\rm down\ type \ quark}:&  d_L\propto h q  \nonumber \\
 &{\rm Higgs \ particle}:&
 \phi\propto \bar h h, \ \  \mbox{$s$-wave} \nonumber \\
 &W^3\! \!- \!{\rm boson}: &   W^3\propto \bar h h,
 \ \ \mbox{$p$-wave} \nonumber \\
 &W^-\! \!- \!{\rm boson}: &   W^- \propto h h, \ \ \mbox{$p$-wave}
 \nonumber \\
 &W^+\! \!-\!{\rm boson}: & W^+ \propto (h h)^\dagger, \ \ \mbox{$p$-wave}. 
\end{eqnarray}

These bound states have to be normalized properly. We shall consider
this issue later on. Using a non-relativistic notation, we can say
that the scalar Higgs particle is a $\bar h h$-state in which the two
constituents are in an $s$-wave.  The $W^3$-boson is the orbital
excitation ($p$-wave). The $W^+$ ($W^-$)-bosons are $p$-waves as well,
composed of $(h h)$ $(\bar h \bar h)$ respectively. Due to the $SU(2)$
structure of the wave function there are no $s$-wave states of the type
$( h h)$ or $(\bar h \bar h)$.

As usual in a quantum field theory, the problem is to identify the
physical degrees of freedom. To do so we have to choose the gauge in
the appropriate way. The Higgs doublet can be used to fix the gauge.
Using the gauge freedom of the local $SU(2)$ group we perform a gauge
rotation such that the scalar doublet takes the form:

\begin{eqnarray}
   h_i=\left ( \begin{array}{c} F+h_{(1)} \\ 0 \end{array}\right),
\end{eqnarray}
where the parameter $F$ is a real number.  If $F$ is sufficiently
large we can perform an $1/F$ expansion for the fields defined above.
We have

\begin{eqnarray} \label{def2}
 \nu_L&=&\frac{1}{F}(\bar h l)=l_1+\frac{1}{F} h_{(1)} l_1\approx l_1 
  \nonumber \\
  e_L&=&\frac{1}{F}(\epsilon^{ij} h_i l_j)=
  l_2+\frac{1}{F} h_{(1)} l_2\approx l_2 
  \nonumber \\
   u_L&=&\frac{1}{F}(\bar h q)=q_1+\frac{1}{F} h_{(1)} q_1\approx q_1 
   \nonumber \\
     d_L&=&\frac{1}{F}(\epsilon^{ij} h_i q_j)=q_2
     +\frac{1}{F} h_{(1)} q_2\approx q_2
     \nonumber \\
     \phi&=&\frac{1}{2 F}(\bar h h)=h_{(1)}+\frac{F}{2} +\frac{1}{2 F} 
     h_{(1)} h_{(1)} \approx h_{(1)}+\frac{F}{2}
     \nonumber \\
     W^3_{\mu}&=& \frac{2 i}{g F^2} ( \bar h D_{\mu} h) = \left ( 1 + 
     \frac{h_{(1)}}{F} \right)^2 B^3_{\mu} + \frac{2 i}{g F} \left (1+ 
     \frac{h_{(1)}}{F} \right) \partial_{\mu} h_{(1)} \approx B^3_{\mu}
    \nonumber \\
   W^-_{\mu}&=& \frac{\sqrt{2} i}{g F^2} ( \epsilon^{ij} h_i D_{\mu} h_j) = \left ( 1 + 
     \frac{h_{(1)}}{F} \right)^2 B^-_{\mu} \approx B^-_{\mu},
    \nonumber \\
   W^+_{\mu}&=& \left (\frac{\sqrt{2} i}{g F^2 } ( \epsilon^{ij} h_i D_{\mu} h_j)\right)^\dagger
 = \left ( 1 + 
     \frac{h_{(1)}}{F} \right)^2 B^+_{\mu} \approx B^+_{\mu},
\end{eqnarray}
where $g$ is the coupling constant of the gauge group $SU(2)_{L}$ and
$D_{\mu}$ is the corresponding covariant derivative.  As can be seen
from (\ref{def2}), the physical particles are those appearing in the
standard model. This is a consequence of the complementarity between
the Higgs phase and the confinement phase. We adopt the usual notation
$B^\pm_\mu=(B^1_\mu \mp i B^2_\mu)/\sqrt{2}$.  The terms which are
suppressed by the large scale $F$ are as irrelevant as the terms which
are neglected in the Higgs phase when the Higgs field is expanded near
its classical vacuum expectation value.  If we match the expansion for
the Higgs field $\phi=h_{(1)}+\frac{F}{2}$ to the standard model, we
see that $F=2 v=492$ GeV where $v$ is the vacuum expectation value.
This parameter can be identified with a typical scale for the theory
in the confinement phase. The physical scale is defined as
$\Lambda=F/\sqrt{2}$, the $\sqrt{2}$ factor is included here because
the physical parameter is not $v$ but $v/\sqrt{2}$ as can be seen from
the Lagrangian (\ref{SML}). We see in the expansion for $W^3_\mu$ that
the suppressed irrelevant terms start at the order $2/F$. We thus
interpret the typical scale for the confinement of $W^3_\mu$ as
$\Lambda_W=\sqrt{2} F/4=173.9$ GeV.

\section{A global SU(2) symmetry}

In the absence of the $U(1)$ gauge group the theory has a global
$SU(2)$ symmetry besides the local $SU(2)$ gauge symmetry. The scalar
fields and their complex conjugates can be written in terms of two
doublets arranged in the following matrix:
\begin{eqnarray}
M=\left(\begin{array}{cc} h_1 & h_2^* \\ h_2 & -h_1^* \end{array}
\right ).
\end{eqnarray}
 The potential of the scalar field $V(h h^*)$ depends solely on
\begin{eqnarray}
 h^* h&=& h_1^* h_1 +h_2^* h_2 \\ \nonumber 
 &=&(\mbox{Re}\, h_1)^2 +  (\mbox{Im\,} h_1)^2
 + (\mbox{Re}\, h_2)^2 +  (\mbox{Im}\, h_2)^2 =- \mbox{det} M.
\end{eqnarray}

This sum is invariant under the group $SO(4)$, acting on the real
vector $(\mbox{Re} \, h_1,\mbox{Im} \, h_1,\mbox{Re} \, h_2,\mbox{Im}
\, h_2)$.  This group is isomorphic to $SU(2) \otimes SU(2)$. One of
these groups can be identified with the confining gauge group
$SU(2)_L$, since $\mbox{det} M$ remains invariant under $SU(2)_L$:
\begin{eqnarray}
  \mbox{det} (UM)= \mbox{det} (M), \  U \in SU(2)_L.
\end{eqnarray}

Now the second $SU(2)$ factor can be identified by considering the matrix
$M^\top$

\begin{eqnarray}
M^\top=\left(\begin{array}{cc} h_1 & h_2 \\ h_2^* & -h_1^* \end{array}
\right ).
\end{eqnarray}
The determinant of $M^\top$, which is equal to $\mbox{det} (M)$,
remains invariant under a $SU(2)$ transformation acting on the
doublets $(h_1,h_2^*)$ and $(h_2,-h_1^*)$.

These transformations commute with the $SU(2)_L$ transformations. They
constitute the flavor group $SU(2)_F$, which is an exact symmetry as
long as no other gauge group besides $SU(2)_L$ is present. With
respect to $SU(2)_F$ the $W$-bosons form a triplet of states
$(W^+,W^-,W^3)$.  The left-handed fermions form $SU(2)_F$ doublets.
Both the triplet as well as the doublets are, of course, $SU(2)_L$
singlets. Once we fix the gauge in the $SU(2)_L$ space such that
$h_2=0$ and $\mbox{Im} \, h_1 =0$, the two $SU(2)$ groups are linked
together, and the $SU(2)_L$ doublets can be identified with the
$SU(2)_F$ doublets. The global and unbroken $SU(2)$ symmetry dictates
that the three $W$-bosons states, forming a $SU(2)_F$ triplet, have
the same mass. Once the Yukawa-type interactions of the fields $e_R$,
$u_R$ and $d_R$ with the corresponding left-handed bound systems are
introduced, the flavor group $SU(2)_F$ is in general explicitly broken.

\section{Electromagnetism and mixing}

The next step is to include the electromagnetic interaction. The gauge
group is $SU(2)_L \otimes U(1)_Y$, where $Y$ stands for the
hypercharge. The covariant derivative is given by

 \begin{eqnarray} \label{covder}
        D_{\mu}=\partial_{\mu}-i \frac{g'}{2} Y {\cal A}_\mu -
        i \frac{g}{2} \tau^a B^a_{\mu}.
       \end{eqnarray}
 The assignment for $Y$ is as follows:
   \begin{eqnarray}
   Y \left( \begin{array}{c}
     l_1\\
   l_2
   \end{array} \right) & = & \left( \begin{array}{cc}
                 1 & 0\\
           0 &  1 \end{array} \right)
         \left( \begin{array}{c}
         l_1\\
         l_2 \end{array} \right) \nonumber \\
   Y \left( \begin{array}{c}
     q_1\\
   q_2
   \end{array} \right) & = & \left( \begin{array}{cc}
                  - \frac{1}{3}& 0\\
           0 & - \frac{1}{3} \end{array} \right)
         \left( \begin{array}{c}
             q_1\\
             q_2 \end{array} \right)\nonumber \\
    Y \left( \begin{array}{c}
     h_1\\
   h_2
   \end{array} \right) & = & \left( \begin{array}{cc}
                     1  & 0 \nonumber \\
              0 &  1 \end{array} \right)
         \left( \begin{array}{c}
             h_1\\
             h_2 \end{array} \right).
\end{eqnarray}
The complete Lagrangian of the model in the confinement phase is given by:
\begin{eqnarray}
  {\cal L}_{c}&=&-\frac{1}{4} G^{a}_{\mu \nu}  G^{a \mu \nu}
                 -\frac{1}{4} f_{\mu \nu}  f^{\mu \nu}
                 + \bar l_L i \fmslash{D} l_L+ \bar q_L i \fmslash{D} q_L
                 +\bar e_R i \fmslash{D} e_R \\ \nonumber &&
                 +\bar u_R i \fmslash{D} u_R
                 + \bar d_R i \fmslash{D} d_R+G_e \bar e_R (\tilde{h} l_L)
                 +G_d \bar d_R (\tilde{h} q_L)\\ \nonumber &&
                 +G_u \bar u_R (h q_L)
                 +h.c.
                 +\frac{1}{2}(D_{\mu} h)^\dagger
                 (D^{\mu} h)-\frac{m^2}{2} h h^\dagger
                 -\frac{\lambda}{4} (h h^\dagger)^2,
\end{eqnarray}
where $m^2>0$ and
\begin{eqnarray} \label{gmunu}
G^a_{\mu\nu}&=&\partial_\mu B^a_\nu-  \partial_\nu
B^a_\mu+g \epsilon^{abc}B^b_\mu B^c_\nu, \\ \nonumber
f_{\mu\nu}&=&\partial_\mu {\cal A}_\nu-  \partial_\nu
{\cal A}_\mu.
\end{eqnarray}

The $U(1)$ gauge group is an unbroken gauge group, like $SU(2)_L$. The
hypercharge of the $h$ field is $+1$, and that of the $h^*$ field is $-1$,
i.e.  the members of the flavor group $SU(2)_F$ have different charge
assignments. Thus the group $SU(2)_F$ is dynamically broken, and a mass
splitting between the charged and neutral vector bosons arises.  The
neutral massive vector boson which is proportional to $(\bar h D_\mu
h)$ and which is not a gauge boson couples to the gauge boson ${\cal
  A}_\mu$, and as a result these bosons are not mass eigenstates, but
mixed states.  The strength of this mixing depends on the internal
structure of the massive bosons.

We shall like to emphasize that the $B^a_\mu$ gauge bosons are as
unphysical as the gluons are in QCD. The hyperphoton ${\cal A}_\mu$ is
not the physical photon $A_\mu$ which is a mixture of ${\cal A}_\mu$
and of the bound state $(\bar h D_\mu h)$. The fundamental D-quarks do
not have an electric charge but only a hypercharge. These hypercharges
give a global hypercharge to the bound states, and one can see easily
that a bound state like the electron has a global hypercharge and will
thus couple to the physical photon, whereas a neutrino has a vanishing
global hypercharge and thus will remain neutral with respect to the
physical photon. So we deduce that QED is not a property of the
microscopic world described by ${\cal L}_{c}$ but rather a property of
the bound states constructed out of these fundamental fields. The
theory in the confinement phase apparently makes no prediction
concerning the strength of the coupling between the bound states and
the electroweak bosons and the physical photon. This information can
only be gained in the Higgs phase.

The mixing between the two states can be studied at the macroscopic
scale, i.e. the theory of bound states, where one has
\begin{eqnarray}
         Z_\mu^0&=&  W^3_\mu \cos{\theta_W} + {\cal A}_\mu 
         \sin{\theta_W}
         \nonumber \\
         A_\mu&=& -   W^3_\mu \sin{\theta_W} + {\cal A}_\mu 
         \cos{\theta_W}.
\end{eqnarray}
Here $\theta_W$ denotes the electroweak mixing angle, and $A_\mu$
denotes the photon field.

\section{The weak mixing angle}
In this section, we want to calculate the typical scale for the
confinement of the $W^3_\mu$-boson. In section \ref{sec2}, we have
matched the expansion for the Higgs field to the standard model. Using
this point of view based on the effective theory concept, we obtained
a scale of $\Lambda_W=173.9$ GeV for this boson. Here we shall
consider a effective Lagrangian to simulate the effect of the
$SU(2)_L$ confinement.

This Lagrangian was originally considered in an attempt to describe
the weak interactions without using a gauge theory \cite{Bjorken}.
This effective Lagrangian is given by
\begin{eqnarray} \label{effLag}
 {\cal L}_{eff}&=&-\frac{1}{4} F_{\mu \nu} F^{\mu \nu}
                            -\frac{1}{4} W^a_{\mu \nu} W^{a \mu \nu}
                            -\frac{1}{2} m^2_W W^a_\mu W^{a \mu}
                            \nonumber \\ &&
                              -\frac{1}{4} \lambda \left
                              ( F_{\mu \nu} W^{3 \mu \nu}
                              + W^3_{\mu \nu} F^{\mu \nu} \right )
                            \nonumber
 \end{eqnarray}
where we have
\begin{eqnarray}
           W^a_{\mu \nu}&=& \partial_\mu W^a_\nu -  \partial_\nu W^a_\mu,
          \nonumber \\ 
          F_{\mu \nu}&=& \partial_\mu {\cal A}_\nu - \partial_\nu {\cal A}_\mu.
\end{eqnarray}

\begin{figure}
\begin{center}
\epsffile{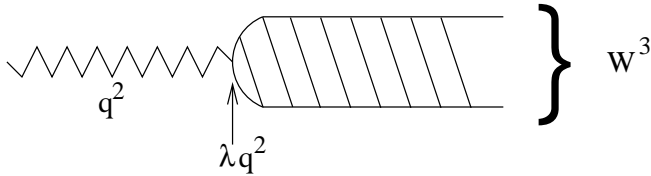}
\caption{Hyperphoton transition into a $W^3$} 
\protect{\label{vertex}}
\end{center}
\end{figure}

The first term in the effective Lagrangian (\ref{effLag}) describes
the field of the hyperphoton, the second term three spin one bosons
and the third term is a mass term which is identical for the three
spin one bosons.  In our case, the fourth term describes an effective
mixing between $W^3$-boson and the hyperphoton.

The effective mixing angle of the Lagrangian given in
equation (\ref{effLag}) reads
\begin{eqnarray}
  \sin^2\theta=\frac{e}{g} \lambda.
  \end{eqnarray}
  
  Using the complementarity principle, we deduce that the mixing angle
  of the theory in the confinement phase has to be the weak mixing angle
  and therefore $\lambda=\sin\theta_W$.
  
  The diagram in figure \ref{vertex} enables us to relate the mixing
  angle to a parameter of the standard model in the confinement phase,
  the typical scale $\Lambda_W$ for the confinement of the
  $W^3$-boson.  For the annihilation of a $W^3$-boson into a
  hyperphoton we have the following relation
  \begin{eqnarray}
    \langle 0 | J^\mu_{Y}(0)|W^3\rangle&=& \frac{\epsilon^\mu}{\sqrt{2 E_W}}
    \frac{m_W^2}{f_W}=\frac{\epsilon^\mu}{\sqrt{2 E_W}} m_W F_W,
\end{eqnarray}
where $J^\mu_{Y}$ is the hyper-current, $F_W=m_W/f_W$ is the decay
constant of the $W^3$-boson, and $\epsilon^\mu$ is its polarization. The
energy of the boson is $E_W$ and the decay constant is defined as follows:
\begin{eqnarray}
    \lambda=\frac{e}{f_W}.
\end{eqnarray}
On the other hand, this matrix element can be expressed using the wave
function of the $W^3$-boson which is a $p$-wave
\begin{eqnarray}
    \langle 0 | J^\mu_{Y}(0)|W^3\rangle&=& \frac{\epsilon^\mu}{\sqrt{2 E_W}}
    \sqrt{\frac{2}{m_W}} \partial_r \phi(0).
\end{eqnarray}
This leads to the following relation for the mixing angle
\begin{eqnarray}
      \sin^2\theta_{W}&=&\frac{8 \pi \alpha}{m_W^5}
      \left(\partial_r \phi(0)\right)^2,
\end{eqnarray}
where $\alpha\approx 1/128$ at $m_W$ is the fine structure constant.

We shall now consider two different models for the wave function:
\begin{itemize}
     \item[a)] Coulombic model.\\
       We adopt the following
       ansatz for the radial wave function
         \begin{eqnarray} 
           \phi(r)&=& \frac{1}{\sqrt{3}} \left(\frac{1}{2 r_B}\right)^{3/2}
           \frac{r}{r_B} \exp\left(-\frac{r}{2 r_B}\right),
           \end{eqnarray}
           where $r_B$ is the Bohr radius. Thus we obtain
    \begin{eqnarray}
      r_B^{-1}&=&m_W\left(\frac{\pi \alpha}{3 \sin^2\theta_W} \
      \right)^{-1/5}.
 \end{eqnarray}
 If we define the typical scale for confinement as $\Lambda_W=
 r_B^{-1}$, we obtain $\Lambda_H= 157$ GeV.
    \item[b)] Three-dimensional harmonic oscillator.
      \\
      The radial part of the wave function is defined as follows:
    \begin{eqnarray}
           \phi({r})&=& \sqrt{\frac{8}{3}} \frac{\beta^{3/2}}{\pi^{1/4}}
           \beta r \exp\left(-\frac{\beta^2 r^2}{2}\right),
\end{eqnarray}
where $\beta=\sqrt{m_W \omega}$, $\omega$ being the frequency of the
oscillator. We identify the typical confinement scale $\Lambda_W$ with
the energy $E=\left(n+\frac{3}{2}\right)\omega$ corresponding to the
quantum number of a $p$-wave i.e. $n=1$, and we obtain
\begin{eqnarray}
            \omega&=&m_W
            \left(\frac{3 \sin^2\theta_W}{64 \alpha \pi^{1/2}}\right)^{2/5}
\end{eqnarray}
and $\Lambda_W=\frac{5}{2} \omega=182$ GeV.
\end{itemize}
Although we have performed a non-relativistic calculation, we see that
the values we find for the typical composite scale are in good
agreement with the naive guess we made based on the concept of an
effective theory.

In order to estimate the value of $\sin^2 \theta_W$, we had to rely on
the simple models, discussed above. However we should like to point
out that $\sin^2 \theta_W$ is not a free parameter in our approach but
fixed by the confinement dynamics. Thus the mixing angle can in
principle be calculated taking e.g. the three dimensional harmonic
oscillator:
\begin{eqnarray} \label{wein}
      \sin^2\theta_{W}&=&\frac{256}{375} \sqrt{10 \pi} \alpha
      \left(\frac{\Lambda_W}{m_W}\right)^{5/2}.
\end{eqnarray}
We can insert the value for $\Lambda_W$ obtained from the effective
theory point of view in equation (\ref{wein}) and we obtain
$\sin^2\theta_W=0.2056$ which has to be compared to the experimental
value $(\sin^2\theta_W)_{exp}=0.23124(24)$. Of course this is a naive
non-relativistic and model dependent calculation. Such a calculation
could also be done by lattice methods but it remains to see whether
this can be carried out in the future.

\section{Deviations from the standard model}
In this section we discuss possible deviations from the standard
model. A possible deviation from the standard model could be the ratio
of the Higgs boson mass to the $W^3$-boson mass. This ratio can take
different values in the Higgs phase and in the confinement phase
\cite{seilerprivate}. Thus the mass of the Higgs boson allows to test
which phase describes the electroweak interactions.

Another aspect is the number of physical states. Obviously the
particle spectrum of the model in the confinement is much richer than
in the standard model since many new $SU(2)_L$ bound states can be
constructed. Especially the $d$-waves, $D^1_{\mu \nu}$ and $D^2_{\mu
  \nu}$ of the bound state of two scalar D-quarks can be constructed.
These bound states have the following expansions:
\begin{eqnarray} \label{def3}
   D^1_{\mu \nu}&=& \frac{2 i}{g F^2} ( \bar h [D_{\mu},D_{\nu}]  h) =
   \left ( 1 + 
     \frac{h_{(1)}}{F} \right)^2 G_{\mu \nu}^3
    \approx G_{\mu \nu}^3
    \\ \nonumber
   D^2_{\mu \nu}&=& \frac{ 2 i}{g F^2}
   ( \epsilon^{ij}h_i [D_{\mu},D_{\nu}]  h_j) =
   \left ( 1 + 
     \frac{h_{(1)}}{F} \right)^2 \left (G_{\mu \nu}^1 + i G_{\mu \nu}^2
     \right)
     \\  \nonumber &&
   \approx \left (G_{\mu \nu}^1 + i G_{\mu \nu}^2
     \right),
\end{eqnarray}
where $G_{\mu \nu}^a$ was defined in equation (\ref{gmunu}).
The masses of such states could in principle be calculated using the
lattice approach or at least related to the mass of the
$W^\pm$-bosons. We also expect the appearance of bound states of the
form $\epsilon^{ij} l_i l_j$, $\bar l l$, $\epsilon^{ij} q_i q_j$ and
$\bar q q$ or the ``mixtures'' of the kind $\bar l q$ and
$\epsilon^{ij} l_i q_j$.  They have the following hypercharges
\begin{eqnarray}
               Y(\epsilon^{ij} l_i l_j )&=& 2, \
               Y(\bar l l)= 0, \ 
               Y(\epsilon^{ij} q_i q_j )= -2/3, \
               Y(\bar q q)= 0, \\ \nonumber
               Y(\epsilon^{ij} l_i q_j )&=& 2/3, \
               Y(\bar l q)= -4/3.
\end{eqnarray}
We cannot assign QED charges to these bound states since they are not
present in the Higgs phase. We expect also the existence of radial
excitations of known particles like a $Z^*$ but basically their
couplings to the rest of the particle spectrum and masses are unknown.
But, we do not expect a sizeable coupling of these excited states to
the fermions. They would be coupled primarily to the $W$-bosons, which
consist of the same scalar D-quarks. A sizeable effect ``beyond the
standard model'' could be observed in the reaction $W^+ + W^- \to Z^*
\to W^+ + W^-$, a reaction which can be studied at the Tevatron and
LHC colliders. The neutral $d$-wave state $D^1_{\mu \nu}$ discussed
above could be found in the same reaction.

A major difference with composite models considered in the past is
that we exclude four fermion interactions, which are setting strong
constraints on composite models.  The new particles, in our approach,
should manifest themselves as radiative corrections to known
processes. We should nevertheless keep in mind that not even the
$Z$-boson contribution to the anomalous magnetic moment of the muon
can be identified \cite{Calmet:1977pu}.

Let us return to the problem of quantum electrodynamics.  As long as
the $SU(2)_L$ confinement takes place we cannot introduce the charge operator
\begin{eqnarray}
  Q&=&1/2(\tau^3 - Y)
  \end{eqnarray}
  since it is not a singlet under $SU(2)_L$. But at very small
  distances, when the confinement scale is irrelevant, this operator
  becomes physical and in principle observable.  We can rewrite the
  covariant derivative (\ref{covder}) using the mass eigenstate
  basis. Since the gauge symmetry is not broken we are not forced to do so,
  but we have the freedom to use this basis.  We have
  
\begin{eqnarray}
     D_\mu&=&\partial_\mu -i \frac{g g'}{2 \sqrt{g^2+g'^2}}  A_\mu
     (\tau^3 - Y ) -
     i \frac{1}{2 \sqrt{2}} g  (\tau^+ B^+_\mu+\tau^- B^-_\mu)
     \\ &&
     -i \frac{1}{2 \sqrt{g^2+g'^2}}  Z_\mu \left(g^2 \tau^3+g'^2 Y \right),
\nonumber
\end{eqnarray}
with
\begin{eqnarray}
  \tau^\pm=1/2(\tau^1\pm i \tau^2).
\end{eqnarray}
Thus the electric charges of the leptonic D-quarks, hadronic D-quarks
and scalar D-quarks are
   \begin{eqnarray} \label{charges}
Q \left( \begin{array}{c}
     l_1\\
   l_2
   \end{array} \right) & = & \left( \begin{array}{cc}
               0 & 0\\
           0 & -1 \end{array} \right)
         \left( \begin{array}{c}
         l_1\\
         l_2 \end{array} \right)  \\
Q \left( \begin{array}{c}
     q_1\\
   q_2
   \end{array} \right) & = & \left( \begin{array}{cc}
                  \frac{2}{3}& 0\\
           0 & -\frac{1}{3} \end{array} \right)
         \left( \begin{array}{c}
             q_1\\
             q_2 \end{array} \right)\nonumber \\
Q \left( \begin{array}{c}
     h_1\\
   h_2
   \end{array} \right) & = & \left( \begin{array}{cc}
                    0  & 0  \\
              0 & -1 \end{array} \right)
         \left( \begin{array}{c}
             h_1\\
             h_2 \end{array} \right).\nonumber
\end{eqnarray}

If one would be able to test the charge of an electron D-quark at very
small distances, we would measure the right QED charge.  We can then
deduce that at low energies it is basically not possible to
distinguish the two phases as the known particles are expected to
behave in the same fashion as in the standard model, and the new
excited bound states cannot be produced because they are too heavy. At
intermediary energies one can expect the new bound states to
contribute through quantum processes but at very small distances, the
bound states will disappear, the D-quarks will become free, and they
will have the usual charges as already discussed. The gauge bosons of
the $SU(2)_L$ gauge group will also be liberated and take over the
role of the electroweak bosons which were bound states.  Especially
the gauge boson $B^3_\mu$ will mix with the hyperphoton, and it will
become impossible to distinguish between the two phases.

A frequent problem with models avoiding the Higgs mechanism are flavor
changing neutral currents. Our model can be extended to three families
by introducing the following doublets in the model
\begin{eqnarray}
  l^{(2)}_i= \left(
  \begin{array}{c}
  l^{(2)}_1 \\ l^{(2)}_2
  \end{array}
\right ), \,
l^{(3)}_i= \left(
  \begin{array}{c}
  l^{(3)}_1 \\ l^{(3)}_2
  \end{array}
\right ), \,
 q^{(2)}_i= \left(
  \begin{array}{c}
  q^{(2)}_1 \\ q^{(2)}_2
  \end{array} \,
\right ) {\rm and} \
q^{(3)}_i= \left(
\begin{array}{c}
  q^{(3)}_1 \\ q^{(3)}_2
  \end{array}
\right ).
 \end{eqnarray}
 We can then identify the leptons and quarks of the second family in
 the following way
\begin{eqnarray}
 \nu_{\mu\, L}&=&\frac{1}{F}(\bar h l^{(2)})=l^{(2)}_1
 +\frac{1}{F} h_{(1)} l^{(2)}_{1}\approx l^{(2)}_{1} 
  \nonumber \\
  \mu_L&=&\frac{1}{F}(\epsilon^{ij} h_i l^{(2)}_j)=l^{(2)}_2+\frac{1}{F} 
h_{(1)} l^{(2)}_2\approx l^{(2)}_2 
  \nonumber \\
   c_L&=&\frac{1}{F}(\bar h q^{(2)})=q^{(2)}_1+\frac{1}{F} h_{(1)} q^{(2)
}_1\approx q^{(2)}_1 
   \nonumber \\
     s_L&=&\frac{1}{F}(\epsilon^{ij} h_i q^{(2)}_j)=q^{(2)}_2
     +\frac{1}{F} h_{(1)} q^{(2)}_2\approx q^{(2)}_2.
     \nonumber \\
    \end{eqnarray}
    The fermions of the third family are
   \begin{eqnarray}
 \nu_{\tau\, L}&=&\frac{1}{F}(\bar h l^{(3)})=l^{(3)}_1
 +\frac{1}{F} h_{(1)} l^{(3)}_{1}\approx l^{(3)}_{1} 
 \nonumber \\
  \tau_L&=&\frac{\sqrt{2}}{F}(\epsilon^{ij} h_i l^{(3)}_j)
  =l^{(3)}_2+\frac{1}{F}
 h_{(1)} l^{(3)}_2\approx l^{(3)}_2 
  \nonumber \\
   t_L&=&\frac{1}{F}(\bar h q^{(3)})=q^{(3)}_1+\frac{1}{F} h_{(1)} q^{(3)
}_1\approx q^{(3)}_1 
   \nonumber \\
     b_L&=&\frac{1}{F}(\epsilon^{ij} h_i q^{(3)}_j)=q^{(3)}_2
     +\frac{1}{F} h_{(1)} q^{(3)}_2\approx q^{(3)}_2.
     \nonumber \\
\end{eqnarray}
The electroweak and Higgs bosons are defined in the same way as
previously.  The universality of the weak interactions is respected as
the three families couple with the same strength to the electroweak
bosons. Due to the complementarity between both phases of the model we
do not expect new physical effects as for example flavor changing
neutral currents.  So this model reproduces all the phenomenological
success of the standard model.

We are aware that the new effects we have discussed in this section
could be forbidden or driven up to very high energies by some yet
unknown selection rules. However this looks unlikely since it does not
appear to be very natural.  Another possibility is that these
non-perturbative effects are also present, but hidden in the standard
model Lagrangian in which case the complementarity would not break
down.

\section{Conclusions}

We have considered a model based on the gauge group $SU(2)_L \otimes
U(1)_Y$. This model represents a viable alternative to the electroweak
standard model. The basic assumption is that physical particles are
singlets under $SU(2)_L$. The model allows a dynamical interpretation
of the weak mixing angle which is related to a typical scale for the
confinement of the $W^3$-boson. This is the main achievement of the
model in the confinement phase. This scale can be calculated by means
of simple wave functions. The result of this naive calculation agrees
with the result expected from the effective theory point of view. In
our approach, the weak mixing angle and therefore also the ratio
$m_Z/m_W$ are calculable.

Some deviations from the standard model could be observed but any
prediction is highly dependent on the details of our confining theory,
whose non-perturbative effects are yet to be investigated.  Lattice
theory should be able to shed some light onto these questions.

If our interpretation is correct, all gauge theories describing the
elementary interactions are based on exact gauge symmetries. This
would restore the asymmetry between QCD and the standard model.
Unless there are some unknown selection rules forbidding their
presence, we expect a rich new spectrum of bound states which could
appear at the scale of a few hundred GeV.  If the confinement phase
describes the electroweak interactions, the desert in the TeV region
could be a blooming desert full of interesting new phenomena.
\section*{Acknowledgements}
The authors are grateful to E. Seiler for stimulating discussions
concerning the question of complementarity. We shall also like to
thank J. Jersak, A. Martin, V. Visnjic and P. Weisz for useful
communications. Furthermore we are indebted to C.P. Korthals-Altes, J.
Pati and G. 't Hooft for useful discussions.

\end{document}